\documentclass[aps,pra,twocolumn,superscriptaddress,longbibliography]{revtex4-2}

\usepackage[T1]{fontenc}
\usepackage[utf8]{inputenc}

\usepackage{amsmath,amssymb,mathtools,bm}
\usepackage{graphicx}
\usepackage{hyperref}
\usepackage{xcolor}
\usepackage{physics}

\usepackage{caption}
\usepackage{subcaption}

\newcommand{\rb}{\bm{r}}
\newcommand{\rbp}{\bm{r}'}
\newcommand{\rbpp}{\bm{r}''}

\newcommand{\Lam}{\Lambda}

\begin{document}

\title{On the question of noise as a resource in quantum computing}

\author{J. Montes}
\email{jmontes.3@alumni.unav.es}
\affiliation{Grupo de Sistemas Complejos,
ETSIME,
Universidad Polit\'ecnica de Madrid, Rios Rosas 21, 28003 Madrid, Spain}

\author{F. Borondo}
\email{f.borondo@uam.es}
\affiliation{Departamento de Qu\'imica, Universidad Aut\'onoma de Madrid, Cantoblanco, 28049 Madrid, Spain}

\author{Gabriel G. Carlo}
\email{g.carlo@conicet.gov.ar}
\affiliation{Comisi\'on Nacional de Energ\'ia At\'omica, CONICET, Departamento de F\'isica, Av.~del Libertador 8250, 1429 Buenos Aires, Argentina}

\date{\today}

\begin{abstract}
Noise is usually regarded as the main obstacle to achieving a scalable quantum advantage, but recent evidence in quantum reservoir computing \cite{Domingo2023TakingAdvantageNoise} suggests that certain channels can, in appropriate regimes, improve performance by enriching the reservoir's effective dynamics. Motivated by this idea we propose a geometric mechanism to explain how non-unital noise applied together with a universal gate set leads to a faster approach to Haar-like distributions of the final states. We find that noise of this kind induces an effective volume expansion on the  manifold of pure states. In order to intuitively understand this we use a minimal $1$ qubit model where we take the amplitude damping channel and combine it with a {\emph renormalization} rule that associates to each resulting mixed state a representative pure state. This composition defines a globally expanding nonlinear map on the space of pure states. We analytically derive the local area expansion factor and identify the global expansion threshold. Finally, we combine amplitude damping with the $G3=\{H,T,\mathrm{CNOT}\}$ universal gate set to show how the approach to Haar-like behavior is faster in an appropriate parameter region. This leads us to propose noise as a possible resource in future quantum algorithms.
\end{abstract}

\maketitle

\section{Introduction}

Recent advances in programmable quantum processors have placed quantum computing at the center of the debate on computational advantages beyond classical reach. In particular, random circuit sampling experiments and their variants have motivated claims of {\emph quantum supremacy} on superconducting and photonic platforms \cite{Arute2019Supremacy,Zhong2019GBS}, as well as theoretical analyses of what such supremacy precisely means and under which assumptions it can be regarded as robust \cite{Harrow2017Supremacy}. These demonstrations, however, have been accompanied by controversy and classical (or hybrid) rebuttals that stress-test the boundary between the quantum and the classical in realistic scenarios \cite{Cho2019IBMDoubt,Liu2021Sunway}.

In the current era of noisy intermediate-scale quantum (NISQ) devices operations are noisy and coherence is limited, which degrades the execution of algorithms and any potential scalable advantage. As a result, a substantial part of the effort has focused on error correction (and its scalability) and on output error mitigation \cite{Zhao2022ReviewPRL,Ugwuishiwu2022ReviewIJSTR,Takagi2022LimitsQEM,GuoYang2022MPO,Ai2023SurfaceCode}.
However, can we identify mechanisms by which noise could help to increase effective quantum randomness?

This question has become especially relevant in quantum learning algorithms that exploit complex dynamics as a resource. A paradigmatic example is quantum reservoir computing (QRC), where a circuit (the reservoir) effectively transforms input information and its observables feed a classical model. There is by now a substantial literature establishing the potential of quantum reservoirs and their processing capabilities  \cite{Mujal2021OpportunitiesQRC,FujiiNakajima2017Harnessing,Nakajima2019Boosting,Kutvonen2020Optimizing,
Chen2020Temporal,MartinezPena2020Capacity,MartinezPena2021DPT,Ghosh2019ReservoirProcessing}.
In this context, it has been observed that not all noise is equivalent: whereas unital channels such as depolarizing or dephasing degrade performance, non-unital channels (in particular, \emph{amplitude damping}) can improve it  \cite{Domingo2023TakingAdvantageNoise}. At least in the realm of QRC noise is not to be suppressed in all cases, it could be used as a resource.

In this work we find that a low amplitude damping noise combined with a paradigmatic example of universal gate sets induces a faster  approach to Haar-like behavior than when just considering the noiseless case. In order to explain this behaviour, we propose an effective  geometric mechanism by which non-unital noise generates a dispersive dynamics on the pure states projective manifold. We develop intuition by means of a one qubit model.

The route to this finding and its explanation is organized as follows. In \textbf{Section II} we introduce the amplitude damping channel and define a two-step procedure (noise + renormalization) that induces an effective nonlinear map $F_\gamma$ on the Bloch sphere. In \textbf{Section III} we characterize expansive and contractive regions. In \textbf{Section IV}, by means of the well established majorization criterion \cite{Carlo1,Carlo2} we show that the approach to Haar-like statistics in a noisy random $G3=\{H,T,\mathrm{CNOT}\}$ circuit is faster than in the noiseless one. Finally, in \textbf{Section V} we discuss the implications of these results and summarize the main conclusions.

\section{Effective action of noise on the Bloch sphere}

\subsection{Amplitude damping for one qubit}
In noisy quantum systems states are generally a mixed $\rho$. The state space of a qubit can be conveniently represented through the Bloch decomposition of the density matrix
\begin{equation}
\rho=\frac{1}{2}\left(\mathbb{I}+\rb\cdot \bm{\sigma}\right),
\label{eq:bloch_rho}
\end{equation}
where $\bm{\sigma}=(\sigma_x,\sigma_y,\sigma_z)$ denotes the vector of Pauli matrices and $\rb\in\mathbb{R}^3$ is the Bloch vector.

For pure states one has $\Tr(\rho^2)=1$, which is equivalent to $\|\rb\|=1$; therefore, pure states form the surface of the Bloch sphere, $S^2$. A standard parametrization of $\rb$ in angular coordinates is
\begin{equation}
\rb(\theta,\phi)=
\begin{pmatrix}
\sin\theta\cos\phi\\
\sin\theta\sin\phi\\
\cos\theta
\end{pmatrix},
\qquad
0\le \theta\le \pi,\ \ 0\le \phi<2\pi,
\label{eq:bloch_vector_prelim}
\end{equation}
where $\theta$ is the polar angle (measured from the north pole, associated with the ground state \(|0\rangle\)) and $\phi$ is the
azimuthal angle. The area element of the sphere is given by
\begin{equation}
\dd A=\sin\theta\,\dd\theta\,\dd\phi,
\label{eq:area_element_prelim}
\end{equation}
and it will later define the local expansion/contraction factor of the map induced by noise and renormalization.

We consider the amplitude damping (AD) channel, which models the relaxation \(|1\rangle\to |0\rangle\) with
probability $\gamma\in[0,1]$. In Kraus representation the channel reads
\begin{equation}
\mathcal{E}_{\gamma}(\rho)=\sum_{k=0,1}E_k\rho E_k^\dagger,
\label{eq:ad_channel}
\end{equation}
with operators
\begin{equation}
E_0=
\begin{pmatrix}
1 & 0\\
0 & \sqrt{1-\gamma}
\end{pmatrix},
\qquad
E_1=
\begin{pmatrix}
0 & \sqrt{\gamma}\\
0 & 0
\end{pmatrix}.
\label{eq:ad_kraus_prelim}
\end{equation}
This channel is non-unital (in general $\mathcal{E}_{\gamma}(\mathbb{I})\neq \mathbb{I}$), reflecting the directionality of the
dissipative process towards the ground state. This property is key to generating an effective dynamics on the states of the system.

In terms of the Bloch vector, the AD channel acts affinely on the Bloch sphere. If
$\rb=(x,y,z)$ corresponds to $\rho$ in \eqref{eq:bloch_rho}, then the state $\rho'=\mathcal{E}_{\gamma}(\rho)$ has Bloch vector
$\rbp=(x',y',z')$ given by
\begin{equation}
x'=\sqrt{1-\gamma}\,x,\qquad
y'=\sqrt{1-\gamma}\,y,\qquad
z'=(1-\gamma)\,z+\gamma.
\label{eq:ad_bloch_prelim}
\end{equation}
AD contracts all components $(x,y,z)$ (inhomogenously) but also introduces a translation in $z$, shifting the
distribution of states toward the north pole. In particular, an initial pure state (\(\|\rb\|=1\)) is mapped in general to a mixed
state with \(\|\rbp\|<1\).

A simple measure of the mixing induced by the channel can be obtained from the purity. For a qubit,
\begin{equation}
\Tr(\rho'^2)=\frac{1+\|\rbp\|^2}{2},
\label{eq:purity_bloch_prelim}
\end{equation}
so that the length $\|\rbp\|$ directly quantifies the loss of purity.

\subsection{Dynamics of the principal state: renormalization map}
When the mixing is small and the Bloch vector satisfies \(\|\rb\|\approx 1\), with very high probability our measurements will be effectively equivalent to measuring the pure state with \(\|\rb\|=1\), known as the principal state. With the evolution the accumulated noise makes this approximation quickly invalid. But it is still useful in order to understand the step by step process by which non-unital noise contributes to disperse the originally pure state vectors. For this reason we introduce the normalization of our Bloch vector. Also, in the general analysis we relax the restriction of weak noise for completeness.

\begin{figure*}[t]
    \centering
    \includegraphics[width=\textwidth]{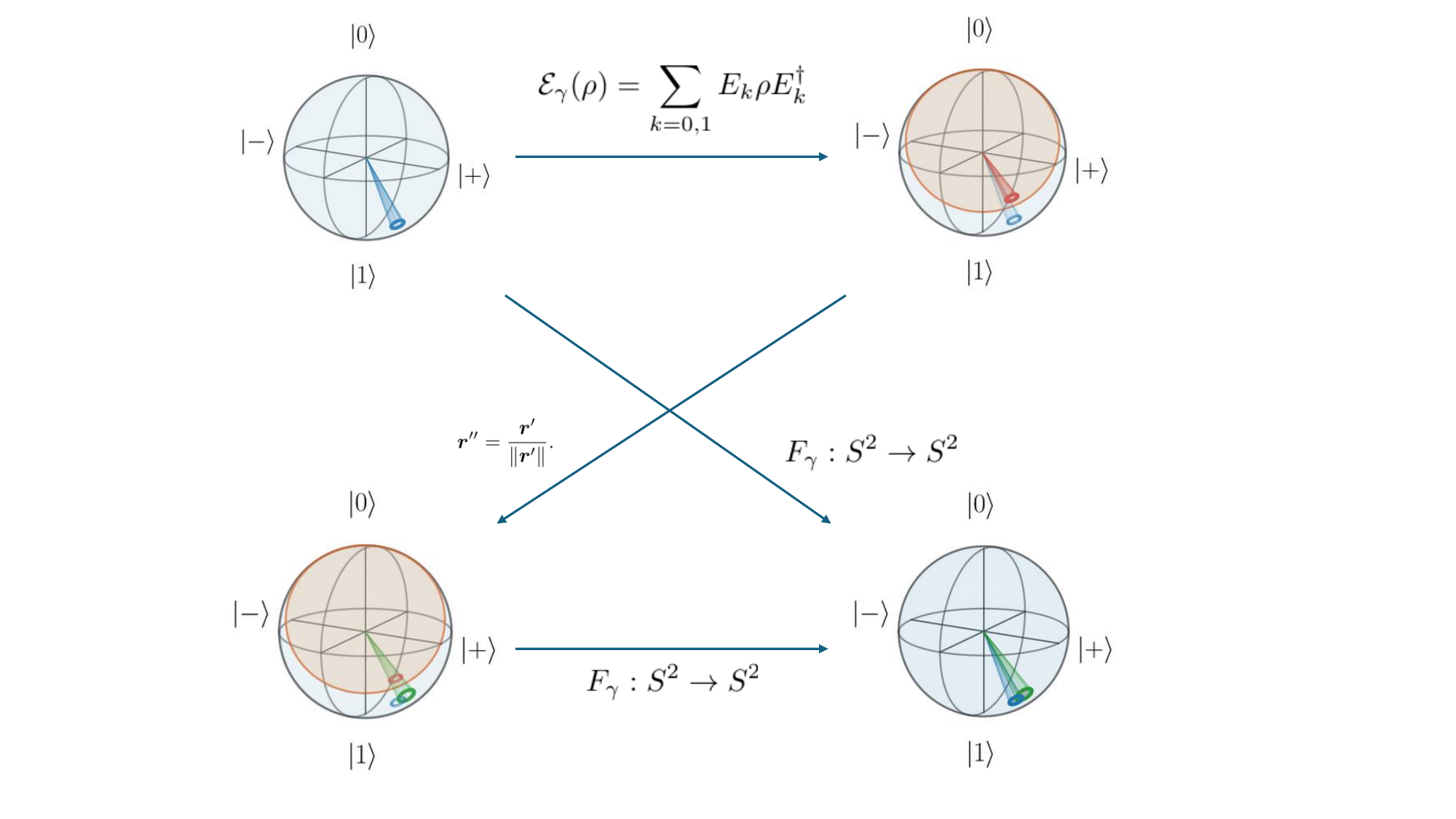}
    \caption{\label{fig:renorm_map}
    Schematic representation of the two-step effective map on the Bloch sphere. Starting from a set of pure states
    (Bloch vectors $\rb\in S^2$), shown in blue, we apply the amplitude damping channel
    $\mathcal{E}_\gamma(\rho)=\sum_{k=0,1}E_k\rho E_k^\dagger$, which maps the states into the interior of the sphere $\rbp$, shown in red.
    Next, we perform the radial projection $\rbpp=\rbp/\|\rbp\|$, obtaining a set of effective pure states, shown in green.
    The composition defines the net nonlinear map $F_\gamma:S^2\to S^2$.}
\end{figure*}

We start from a pure one-qubit state, represented by a point on the Bloch sphere (\(\|\rb\|=1\)). As mentioned, after the action of the
amplitude damping channel, the state generally becomes mixed and its Bloch vector \(\rbp\) lies in the interior of the sphere
(\(\|\rbp\|<1\)). In order to identify a geometric mechanism of effective dispersion over the set of pure (principal) states,
we introduce the following two-step map:
\begin{enumerate}
\item Apply the AD channel, \(\rho \mapsto \rho'=\mathcal{E}_\gamma(\rho)\), equivalently \(\rb \mapsto \rbp\) via
\eqref{eq:ad_bloch_prelim}.
\item Associate to \(\rho'\) an effective pure state by radial projection onto the surface,
\begin{equation}
\rbpp=\frac{\rbp}{\|\rbp\|}.
\label{eq:radial_projection}
\end{equation}
\end{enumerate}
The result is a deterministic nonlinear map on the sphere,
\begin{equation}
F_\gamma: S^2 \to S^2,
\qquad
\rb \longmapsto \rbpp=\frac{\rbp}{\|\rbp\|},
\label{eq:F_gamma_def}
\end{equation}
see Fig.~\ref{fig:renorm_map}.

\subsection{Deformation of the Bloch sphere}

The AD channel by itself acts linearly on the convex set of states (the Bloch ball), and can be understood as an anisotropic
contraction in \((x,y,z)\) combined with a translation in \(z\) \cite{NielsenChuang}. However, to describe when and where dissipative noise can increase the dispersion of an ensemble of initially pure states, it is convenient to separate two conceptually distinct
effects:
(i) the generation of mixing (loss of purity), and
(ii) the directional redistribution of the
{\emph most representative} pure states that emerge from the resulting mixed state.

For a qubit, the radial projection \eqref{eq:radial_projection} selects the point on the sphere aligned with \(\rbp\), i.e.,
it preserves the direction of the state in Bloch space. This choice has two properties that are relevant for our analysis.
On one hand, for weak noise \(\gamma\to 0\) one has \(\rbp\to \rb\) and therefore \(\rbpp\to \rb\), so that \(F_\gamma\) tends to the identity on \(S^2\). On the other, the mixing induced by the channel is encoded in the norm \(N(\theta)=\|\rbp\|\), while the degree of dispersion on the sphere depends on the normalized direction \(\rbpp\). This allows for an easy comparison between the two effects. \(F_\gamma\) acts as an effective geometric map that captures the tendency of noise to deform and redistribute a set of pure states when one focuses on their directional component.

We apply \eqref{eq:ad_bloch_prelim} to the pure state \eqref{eq:bloch_vector_prelim}. Defining \(A:=1-\gamma\), we obtain
\begin{equation}
\rbp(\theta,\phi)=
\begin{pmatrix}
\sqrt{A}\,\sin\theta\cos\phi\\
\sqrt{A}\,\sin\theta\sin\phi\\
A\cos\theta+\gamma
\end{pmatrix},
\label{eq:rprime_theta_phi}
\end{equation}
The intermediate norm \(N(\theta)=\|\rbp\|\) is
\begin{align}
N(\theta)^2
&=A\sin^2\theta+\big(A\cos\theta+\gamma\big)^2 \nonumber\\
&=1-\gamma(1-\gamma)\,(1-\cos\theta)^2,
\label{eq:Ntheta_closed}
\end{align}
and therefore
\begin{equation}
N(\theta)=\sqrt{1-\gamma(1-\gamma)\,(1-\cos\theta)^2}.
\label{eq:Ntheta_final}
\end{equation}
After the radial projection \eqref{eq:radial_projection}, the direction in the \(xy\) plane is preserved, so the azimuthal angle does
not change,
\begin{equation}
\phi'=\phi,
\label{eq:phi_map_section3}
\end{equation}
while the polar angle \(\theta'\) is determined by the \(z\) component of the normalized vector \(\rbpp\),
\begin{equation}
\cos\theta'=\frac{A\cos\theta+\gamma}{N(\theta)}
=
\frac{(1-\gamma)\cos\theta+\gamma}
{\sqrt{1-\gamma(1-\gamma)\,(1-\cos\theta)^2}}.
\label{eq:theta_map_section3}
\end{equation}
Consequently, the map \(F_\gamma\) is axially symmetric and its dynamics reduces to a one-dimensional transformation
\(\theta\mapsto\theta'(\theta)\), with \(\phi\) as a passive coordinate.

From Eq. \eqref{eq:theta_map_section3} becomes clear that the affine term \(A\cos\theta+\gamma\) shifts the \(z\) component towards more positive values (without the renormalization procedure it would contract the whole sphere to the north pole). By reinserting the states into \(S^2\),
local area expansion arises in certain regions since it amounts
to mapping lower measure sections to higher measure ones (effectively increases dispersion on \(S^2\) of the corresponding initial distributions). In what follows, we will refer to this effect as {\emph geometric dispersion}.

\section{Expansive and contractive regions}

The expression for the local area expansion factor \(\Lam(\theta;\gamma)\) is the central ingredient for identifying expansive and
contractive regions of the proposed map. Below we show how it is obtained from the Jacobian of the angular map.

The renormalized map \(F_\gamma:S^2\to S^2\) is given by \(\phi'=\phi\) and by the relation
\begin{equation}
\begin{aligned}
\cos\theta' &= \frac{(1-\gamma)\cos\theta+\gamma}{N(\theta)},\\
N(\theta) &= \sqrt{1-\gamma(1-\gamma)\,(1-\cos\theta)^2}.
\end{aligned}
\label{eq:theta_map_and_N}
\end{equation}

As a consequence, the Jacobian in coordinates \((\theta,\phi)\) has a diagonal structure,
\begin{equation}
J(\theta,\phi)=
\frac{\partial(\theta',\phi')}{\partial(\theta,\phi)}
=
\begin{pmatrix}
\dfrac{\dd\theta'}{\dd\theta} & 0\\[6pt]
0 & 1
\end{pmatrix},
\qquad
\det J(\theta)=\frac{\dd\theta'}{\dd\theta}.
\label{eq:jacobian_det_mainbody}
\end{equation}

The area element on the sphere transforms as \(\dd A'=\sin\theta'\,\dd\theta'\,\dd\phi'\) and \(\dd A=\sin\theta\,\dd\theta\,\dd\phi\).
Therefore, the local area expansion factor is defined as
\begin{equation}
\Lam(\theta;\gamma)
:=\frac{\dd A'}{\dd A}
=\frac{\sin\theta'}{\sin\theta}\,
\left|\frac{\dd\theta'}{\dd\theta}\right|.
\label{eq:Lambda_def_mainbody}
\end{equation}

To evaluate \eqref{eq:Lambda_def_mainbody} explicitly, we first observe that the transverse component of the Bloch vector after AD satisfies
\begin{equation}
x'^2+y'^2=(1-\gamma)\sin^2\theta,
\label{eq:xyprime}
\end{equation}
while after the radial renormalization \(\rbpp=\rbp/N(\theta)\) one has
\begin{equation}
\sin\theta'=\sqrt{(x'' )^2+(y'' )^2}=\frac{\sqrt{x'^2+y'^2}}{N(\theta)}
=\frac{\sqrt{1-\gamma}\,\sin\theta}{N(\theta)}.
\label{eq:sin_ratio}
\end{equation}
Equation \eqref{eq:sin_ratio} immediately yields
\begin{equation}
\frac{\sin\theta'}{\sin\theta}=\frac{\sqrt{1-\gamma}}{N(\theta)}.
\label{eq:sin_ratio_compact}
\end{equation}

The direct calculation of \(\dd\theta'/\dd\theta\) from \eqref{eq:theta_map_and_N} is algebraically nontrivial; for completeness, the
detailed derivation is presented in Appendix~\ref{app:jacobian_theta}. The result can be written in compact form as
\begin{equation}
\frac{\dd\theta'}{\dd\theta}
=
\sqrt{1-\gamma}\,
\frac{1-\gamma\,(1-\cos\theta)}{N(\theta)^2},
\label{eq:dtheta_result_mainbody}
\end{equation}
where the sign is determined by the orientation of the map (in our area analysis we will use the absolute value).

Substituting \eqref{eq:sin_ratio_compact} and \eqref{eq:dtheta_result_mainbody} into \eqref{eq:Lambda_def_mainbody} finally gives
\begin{equation}
\Lam(\theta;\gamma)
=
(1-\gamma)\,
\frac{\left|1-\gamma(1-\cos\theta)\right|}
{\left[1-\gamma(1-\gamma)(1-\cos\theta)^2\right]^{3/2}},
\label{eq:Lambda_closed_mainbody}
\end{equation}
where we used \(N(\theta)^2=1-\gamma(1-\gamma)(1-\cos\theta)^2\).
The condition \(\Lam(\theta;\gamma)>1\) identifies a locally expansive region, while \(\Lam(\theta;\gamma)<1\) corresponds to local
area contraction.

We therefore define the boundary \(\theta_c(\gamma)\in(0,\pi)\) as the value (when it exists) satisfying
\begin{equation}
\Lam\!\left(\theta_c(\gamma);\gamma\right)=1,
\label{eq:theta_c_def_main}
\end{equation}
so that the regions \(\theta<\theta_c(\gamma)\) and \(\theta>\theta_c(\gamma)\) correspond, respectively, to local contraction and
local area expansion. Equivalently, the expansive set for a given \(\gamma\) is
\begin{equation}
\mathcal{E}_\gamma=\left\{(\theta,\phi)\in S^2:\Lam(\theta;\gamma)>1\right\}
=\left\{(\theta,\phi):\theta\in(\theta_c(\gamma),\pi)\right\},
\label{eq:expansive_set}
\end{equation}
provided the solution \(\theta_c(\gamma)\) exists. In what follows we will refer to the region \(\theta\approx\pi\) as the
\emph{neighborhood of the south pole} (states with larger excited-state population), and to \(\theta\approx 0\) as the
\emph{neighborhood of the north pole} (the ground state).

A first asymptotic analysis is in order here. We see the behavior of \(\Lam\) at the poles. Using \eqref{eq:Lambda_closed_mainbody}:

\paragraph*{North pole (\(\theta=0\)).}
Since \(1-\cos\theta=0\) and \(\theta'=\theta\) at the north pole, we obtain
\begin{equation}
\Lam(0;\gamma)=1-\gamma.
\label{eq:Lambda_north}
\end{equation}
Therefore, for any \(\gamma>0\) the map is contractive in a neighborhood of the north pole.

\paragraph*{South pole (\(\theta=\pi\)).}
Since \(1-\cos\pi=2\), the denominator of \eqref{eq:Lambda_closed_mainbody} simplifies via
\(
1-4\gamma(1-\gamma)=(1-2\gamma)^2
\),
and we obtain
\begin{equation}
\Lam(\pi;\gamma)=\frac{1-\gamma}{|1-2\gamma|^2}.
\label{eq:Lambda_south}
\end{equation}
This expression exhibits a singular behavior as \(\gamma\to 1/2\), associated with the fact that the intermediate mixed state
collapses to \(\rbp=\bm{0}\) when starting from the south pole with \(\gamma=1/2\), making the radial projection
\eqref{eq:radial_projection} ill-defined at that isolated point. On the rest of the sphere, and for \(\gamma\neq 1/2\), the map
remains well defined.

Equations \eqref{eq:Lambda_north}--\eqref{eq:Lambda_south} capture the essential asymmetry of the map: while the north pole is always
contractive for \(\gamma>0\), the south pole can be strongly expansive for certain values of \(\gamma\).

The expansive set \(\mathcal{E}_\gamma\) exists if and only if \(\Lam(\theta;\gamma)\) exceeds the threshold \(1\) for some
\(\theta\in(0,\pi)\). Since for $\gamma \to 1$ the expansion reduces to the neighborhood of the south pole, a necessary and sufficient
condition is
\begin{equation}
\Lam(\pi;\gamma)>1.
\label{eq:expansion_condition_south}
\end{equation}
Using \eqref{eq:Lambda_south}, the inequality \(\Lam(\pi;\gamma)>1\) is equivalent to
\begin{equation}
\frac{1-\gamma}{|1-2\gamma|^2}>1
\quad \Longleftrightarrow \quad
\gamma(3-4\gamma)>0.
\label{eq:gamma_threshold_steps}
\end{equation}
Therefore, for \(\gamma\in(0,1)\) one obtains the global threshold
\begin{equation}
0<\gamma<\frac{3}{4}
\quad \Longleftrightarrow \quad
\exists\,\theta:\Lam(\theta;\gamma)>1,
\label{eq:gamma_threshold}
\end{equation}
whereas
\begin{equation}
\gamma\ge \frac{3}{4}
\quad \Longrightarrow \quad
\Lam(\theta;\gamma)\le 1 \ \ \forall\,\theta\in[0,\pi].
\label{eq:no_expansion_region}
\end{equation}
In other words: for sufficiently strong noise (\(\gamma\ge 3/4\)) the renormalized dynamics is globally area-contracting, whereas for
low to intermediate noise (\(0<\gamma<3/4\)) an expansive belt emerges around the south pole.

\begin{figure*}[t]
    \includegraphics[width=\textwidth]{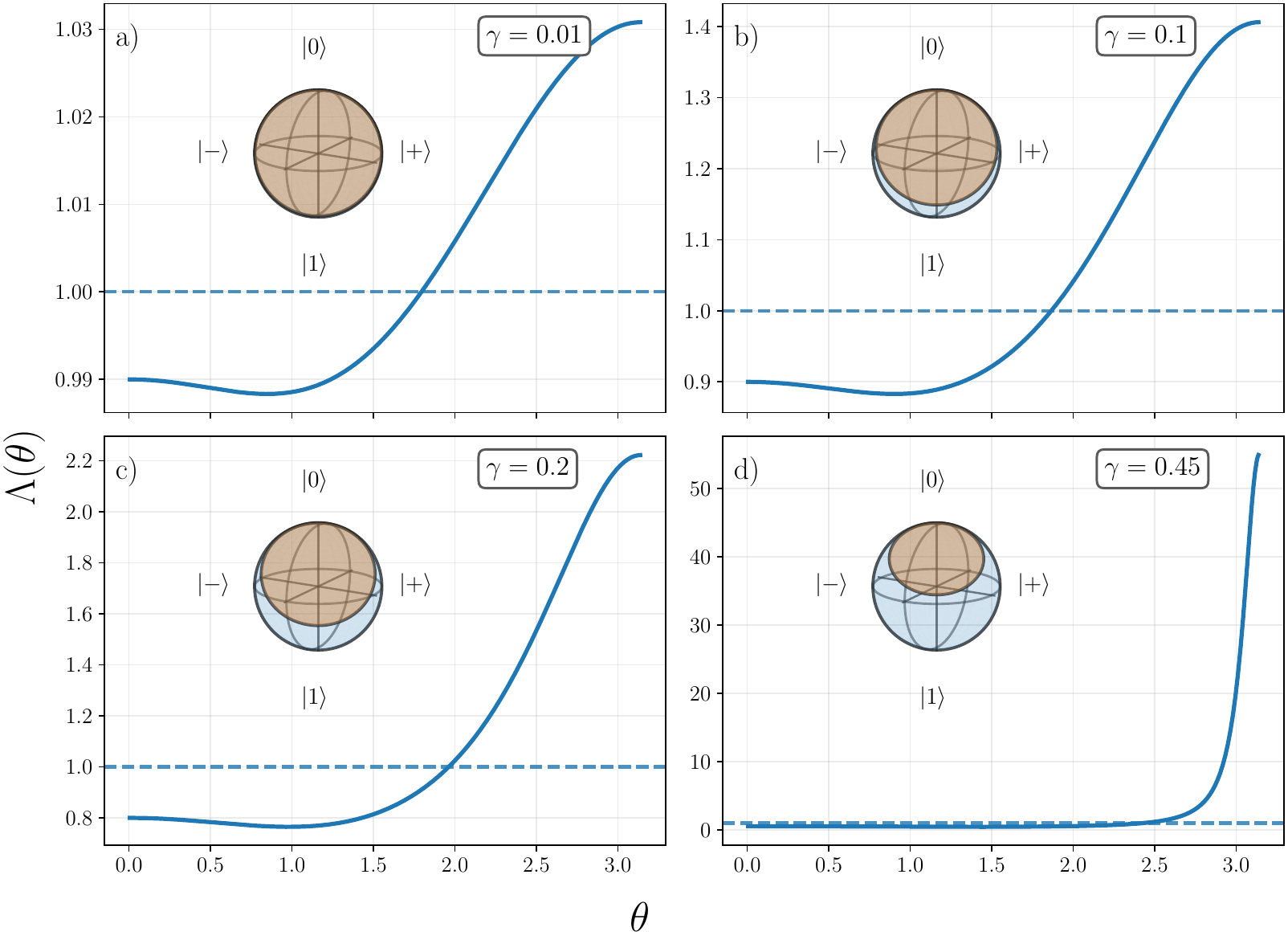}
    \caption{\label{fig:gamma_sweep}
Local area expansion factor $\Lam(\theta;\gamma)$ of the map $F_\gamma$ as a function of the polar angle $\theta$. Panels (a)--(d) correspond to $\gamma=0.01$, $0.1$, $0.2$ and $0.45$, respectively. The
dashed line marks the threshold $\Lam=1$ separating contractive regions ($\Lam<1$) from expansive regions ($\Lam>1$).
\emph{Inset:} Bloch-sphere representation of the effect of the amplitude damping channel prior to renormalization: the outer sphere
corresponds to the set of initial pure states ($\|\rb\|=1$), while the inner sphere shows the image under AD, $\rb\mapsto\rbp$.}
\end{figure*}

Figure~\ref{fig:gamma_sweep} illustrates how the renormalized map produces a trade-off between geometric dispersion and mixing.
For low to intermediate values of $\gamma$ (panels a--d: $\gamma=0.01,\,0.1,\,0.2,\,0.45$) we apply the map $F_\gamma$ and plot the
local area expansion factor $\Lam(\theta;\gamma)$ together with the expansive region (where $\Lam>1$). When $\gamma$ is small
($\gamma=0.01$), the expansive region is maximal in angular extent, but the expansion is very weak, with $\Lam(\theta)$ barely above
unity. For $\gamma=0.1$ the expansive region remains large and the magnitude of the expansion increases appreciably (typical values
around $\Lam\simeq 1.4$). Increasing to $\gamma=0.2$, the expansive area shrinks, but the local expansion grows strongly; however, as
shown in the \emph{inset}, the degree of mixing is already significant (loss of purity associated with $N(\theta)=\|\rbp\|$). Finally, for
$\gamma=0.45$ the expansion is large but becomes confined to a very small neighborhood of the south pole, and the purity is low.

For \(\gamma\in(0,3/4)\), Eq.~\eqref{eq:theta_c_def_main} defines a boundary \(\theta_c(\gamma)\) separating the contractive region
(near the north pole) from the expansive region (near the south pole). Although \(\theta_c(\gamma)\) does not in general admit a
simple closed form, it is obtained directly as the one-dimensional solution of \(\Lam(\theta;\gamma)=1\) and can be
numerically calculated.

\begin{figure*}[t]
    \centering
    \includegraphics[width=\textwidth,height=0.85\textheight,keepaspectratio]{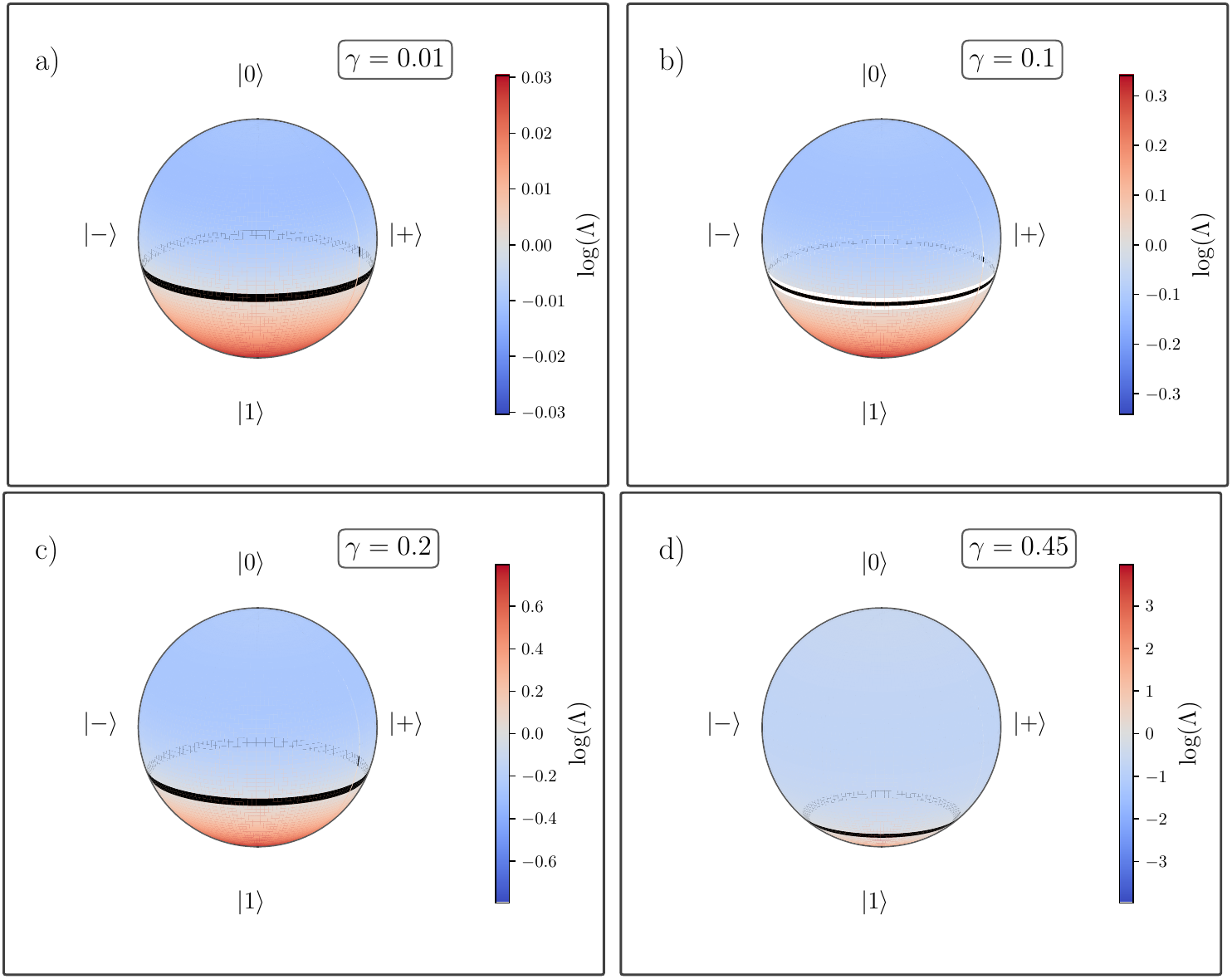}
    \caption{\label{fig:bloch_loglambda}
    Map of the local area expansion factor over the Bloch sphere, represented as
    $\log\!\big(\Lam(\theta;\gamma)\big)$. Blue tones indicate contractive regions
    ($\Lam<1$) and red tones expansive regions ($\Lam>1$). The black curve marks the boundary $\Lam(\theta_c(\gamma);\gamma)=1$ that
    separates both dynamics. Panels (a)--(d) correspond to $\gamma=0.01$, $0.1$, $0.2$ and $0.45$, respectively.}
\end{figure*}

The local analysis in terms of \(\Lam(\theta;\gamma)\) identifies which regions of the sphere are expansive or contractive. Though we are interested in the very weak noise limit, it is interesting to visualize the general behaviour of the renormalization map.
Figure~\ref{fig:bloch_loglambda} provides a global idea of the expansion/contraction induced by $F_\gamma$ on $S^2$ in terms of $\log(\Lam)$. Consistently with Fig.~\ref{fig:gamma_sweep}, for $\gamma=0.01$ (panel a) the expansive region appears as a
broad belt around the south pole, but with very weak intensity (values of $\log(\Lam)$ close to zero). Moving to $\gamma=0.1$ (panel
b) the expansive area remains extensive and the intensity increases appreciably. For $\gamma=0.2$ (panel c) the red zone becomes more
concentrated near the south pole: the expansive area decreases, although the local expansion is stronger. Finally, for $\gamma=0.45$ (panel d) the extreme expansion is confined to a very small neighborhood of the south pole, with the
rest of the sphere dominated by contraction, reflecting the trade-off between expansion intensity and angular extent of the expansive
region.

\section{Noise accelerates approach to Haar-like behaviour}
\label{sec:noise_haar_sdl}

In the previous sections we used a geometric analysis to
show that the same noise channel that generates mixing induces an effective dispersion on the pure-state projective manifold. Here we connect this mechanism with a standard benchmark of random-circuit complexity: the majorization criterion of \cite{Carlo1,Carlo2}. We must underline that our aim is not measuring how noise spoils the quality of a Haar operators ensemble like for example in t-design studies \cite{tDesign}. On the contrary, we are interested in showing how relatively weak noise can lead to an initially faster effective exploration of it. For this purpose the majorization criterion is suitable.

If we have an $n$-qubit pure state $\ket{\psi}\in\mathbb{C}^{d}$ with $d=2^n$, we consider computational-basis measurement probabilities
\begin{equation}
P(i)=|\psi_i|^2,\qquad i=1,\dots,d.
\end{equation}
For a mixed state $\rho$ we consider the main diagonal as the probability vector $P(i)$.
Let $P^{\downarrow}$ be the sorted vector in descending order,
$P^{\downarrow}(1)\ge\cdots\ge P^{\downarrow}(d)$, and define the cumulative sums (cumulants)
\begin{equation}
C(k)=\sum_{j=1}^{k}P^{\downarrow}(j),\qquad k=1,\dots,d.
\label{eq:sdl_cum_def}
\end{equation}
Given an ensemble of $M$ states at a given depth $t$, we define the SDL signature as the
sample standard deviation of the cumulants across the ensemble,
\begin{equation}
\mathrm{SDL}(k;t)
=
\sqrt{
\mathbb{E}_m\!\left[C_t^{(m)}(k)^2\right]
-\left(\mathbb{E}_m\!\left[C_t^{(m)}(k)\right]\right)^2}.
\label{eq:sdl_signature}
\end{equation}
We then define a scalar distance to Haar by comparing with a Haar reference $\mathrm{SDL}_{\mathrm{Haar}}$ (estimated from
$M_{\mathrm{Haar}}$ Haar states in dimension $d$):
\begin{equation}
D(t)=\left\|\mathrm{SDL}(t)-\mathrm{SDL}_{\mathrm{Haar}}\right\|_2.
\label{eq:sdl_distance}
\end{equation}
This is the majorization criterion which provides with a compact and efficient diagnostics of Haar-likeness for ensembles, including weak noise scenarios.

We consider random circuits built from the universal gate set $G3=\{H,T,\mathrm{CNOT}\}$.
At each layer, a single gate is drawn with probabilities
\begin{equation}
\mathbb{P}(H)=\mathbb{P}(T)=\mathbb{P}(\mathrm{CNOT})=\frac{1}{3},
\end{equation}
where $H$ or $T$ acts on a uniformly random qubit, and $\mathrm{CNOT}$ acts on a uniformly random ordered pair of distinct qubits.
Starting from $\ket{0}^{\otimes n}$ for instance, we propagate an ensemble of $M$ samples up to depth $T$ and evaluate $D(t)$ at each depth.

We compare two evolutions, the pure state evolution under the random unitary circuit $\ket{\psi}\mapsto U\ket{\psi}$ and the  noisy one (along each unitary gate we apply local amplitude damping on the idle qubits -- we assume perfect gates -- with uniform $\gamma$ values). To reduce Monte Carlo variance in the comparison the same random gate realization is applied in both cases (unitary and noisy). This allows to better see the amplitude damping induced geometric dispersion contribution. We have considered the following simulation parameters:  $M=3000$, $M_{\mathrm{Haar}}=3000$, depths up to $T=200,220$, and $n=6,7$ qubits.

\begin{figure*}[t]
    \centering
    \includegraphics[width=\textwidth]{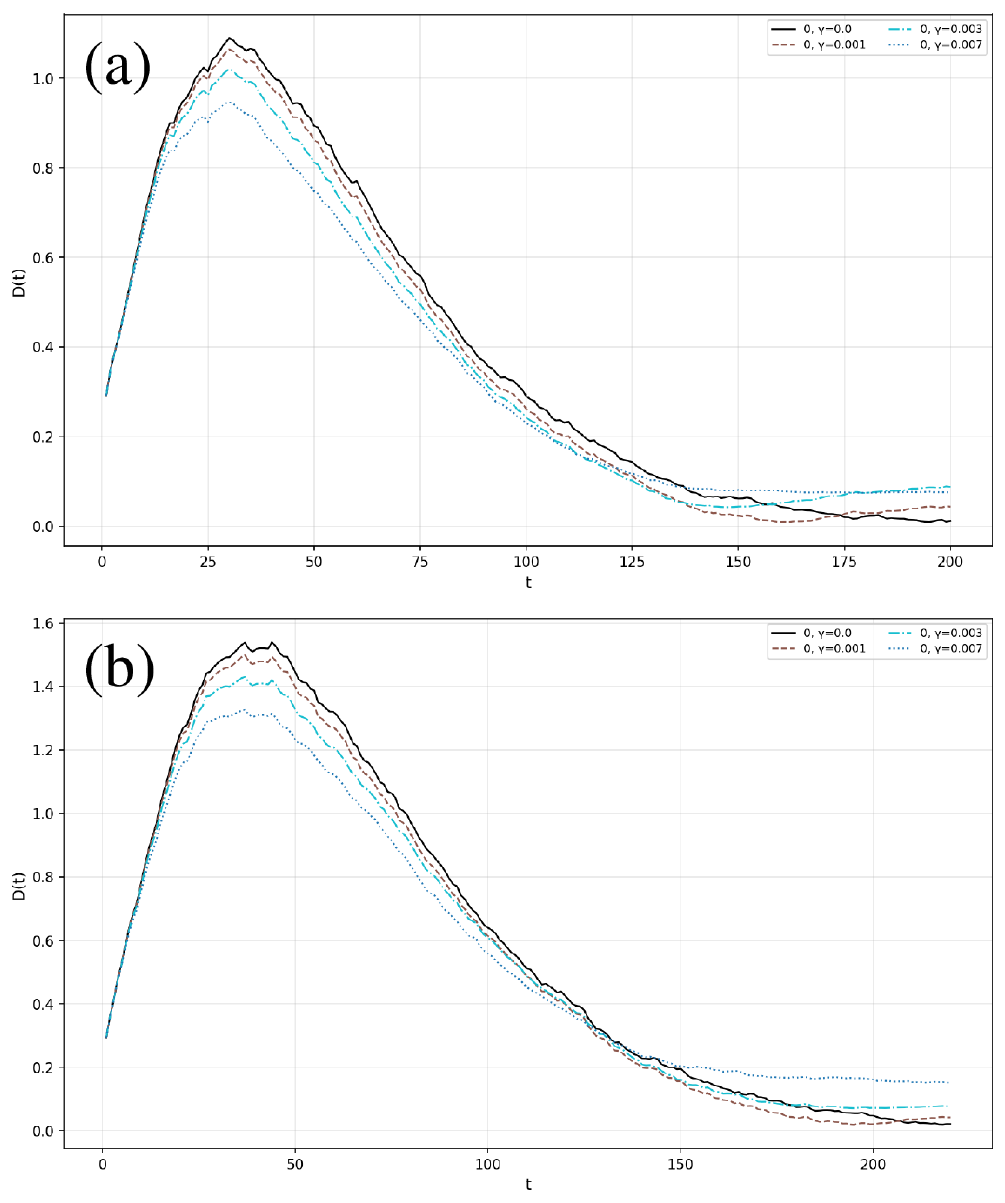}
    \caption{\label{fig:sdl_haar_distance1}
    (a) SDL-distance to Haar,
    $D(t)=\|\mathrm{SDL}(t)-\mathrm{SDL}_{\mathrm{Haar}}\|_2$,
    as a function of circuit depth for random G3 circuits of $n=6$ qubits.
    Solid black line: unitary G3 evolution. Other patterns and colors indicated in the panel: G3 followed by local amplitude damping with different $\gamma$ parameters.
    (b) Same as in panel (a) but for $n=7$ qubits.
    }
\end{figure*}

For $n=6$, we find that the distance to Haar of the diagonal of the density matrix $\rho'$ for $\mathrm{SDL}(t)$ in Figure~\ref{fig:sdl_haar_distance1}(a) reduces faster with noise than with purely unitary dynamics. For weak noise ($\gamma=0.001$), at around $150$ gates the distance to Haar is similar to that of $200$ noiseless gates.  Stronger noise ultimately introduces a bias that limits asymptotic proximity to Haar.
This prevents from an exact coincidence of the cumulant fluctuations with those of Haar. In any case, we underline that the distance $D(t)$ decreases more rapidly than
in the purely unitary case over an extended range of depths.  Operationally, this means that at a fixed finite depth -- a relevant regime in NISQ and reservoir-like settings -- the noisy ensemble can be closer to Haar sooner.
\begin{figure*}[t]
    \centering
    \includegraphics[width=\textwidth]{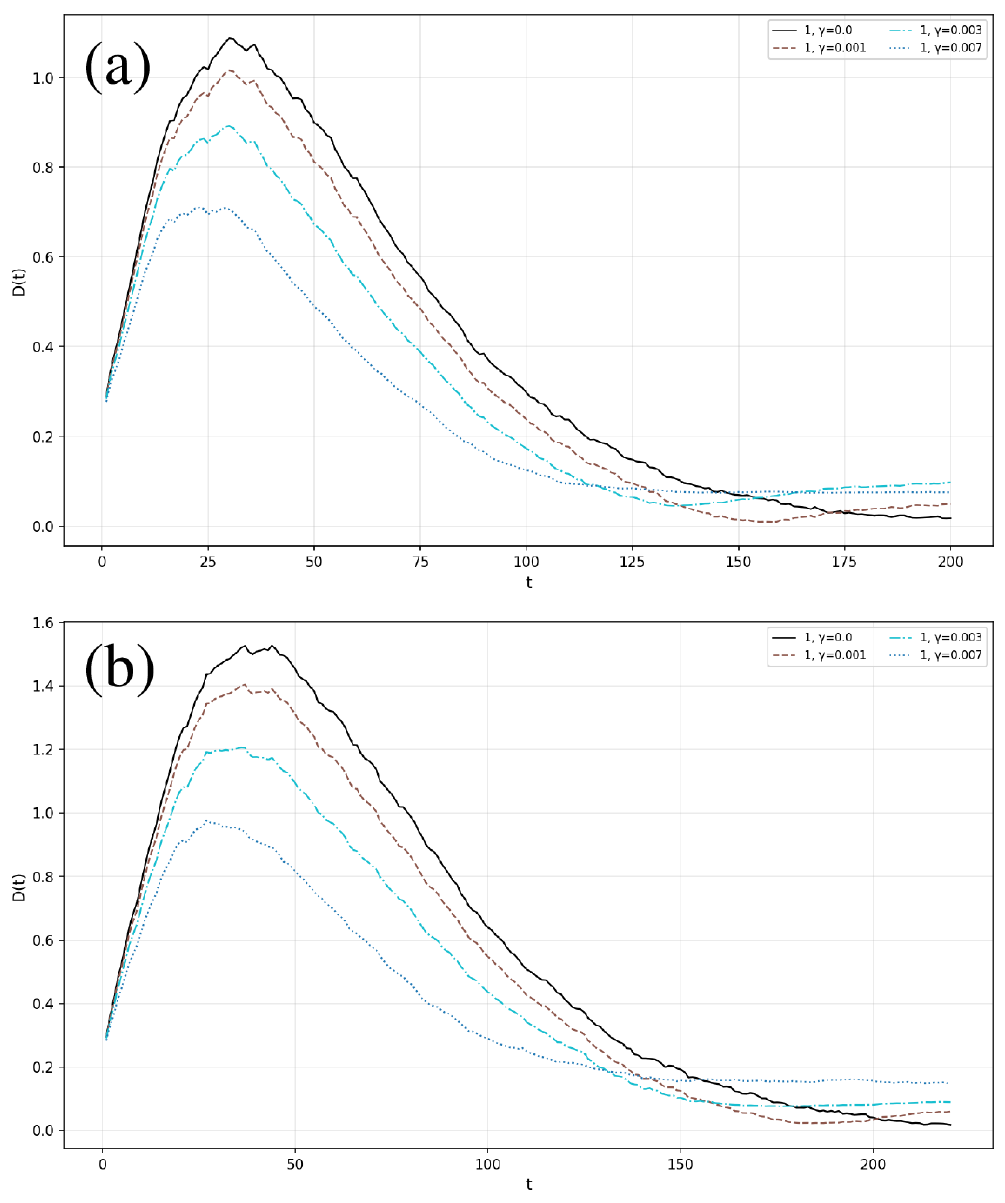}
    \caption{\label{fig:sdl_haar_distance2}
    Same as in Fig. \ref{fig:sdl_haar_distance1}, but considering $\ket{1}^{\otimes n}$ as the initial state.
    }
\end{figure*}

The same behaviour is observed when considering $n=7$ qubits.  Now, at around $180$ gates the distance to Haar is similar to that of $220$ noiseless gates. This can be seen in Figure~\ref{fig:sdl_haar_distance1}(b).

This behavior is naturally understood thanks to our geometric analysis.  The non unital channel effective map on the projective manifold of pure states introduces a directional deformation towards regions of greater measure, globally.  In the weak noise regime, this induced expansion contributes an additional effective dispersion of
states that can enhance the quantum randomness of measurement probabilities at finite depth, thereby reducing $D(t)$ faster. This is further confirmed by looking at Fig. \ref{fig:sdl_haar_distance2} where we have repeated the same simulations leading to Fig.  \ref{fig:sdl_haar_distance1} but initializing the quantum state at $\ket{1}^{\otimes n}$. The approach to haar like behavior is even faster. In fact, this initial state is in the maximal expansion region of each qubit (see Fig. \ref{fig:bloch_loglambda}).

Regarding limitations of this mechanism, these results are in line with the analysis of the previous Sections. As $\gamma$ increases, dissipation becomes dominant and introduces a progressively stronger bias toward the computational zero state. While this can produce a rapid initial decay, it also leads to a restriction with respect to asymptotic relaxation to Haar: for sufficiently large $\gamma$ and/or gate number,
the distance $D(t)$ does not continue decreasing as in the unitary dynamics. Moreover, the approximation to truly Haar-like fluctuations is limited. Non-unital noise provides a finite-depth speedup toward Haar-like statistics, but excessive (eventually accumulated) noise ultimately prevents arbitrarily close convergence by injecting a preferred direction in state space.

\section{Conclusions}
In this work we have shown that non-unital noise can be interpreted as a generator of an effective dynamics when its action is viewed through the dominant pure-state. For the minimal one-qubit case,
the composition of amplitude damping with renormalization induces a nonlinear map $F_\gamma$ on $S^2$ whose geometry is
quantified by the local factor $\Lambda(\theta;\gamma)$. This analysis reveals a sharp separation between contractive zones (near the
north pole) and an expansive belt (near the south pole) that exists only for $0<\gamma<3/4$. When combined with random circuits drawn from a paradigmatic universal gate set like G3 the global effect is an enhanced Haar convergence due to induced geometric dispersion.

We can explain this remarkable result by means of our geometrical analysis of the projection on pure principal states. These
findings complement previous observations in quantum reservoir computing where noise acts as a resource under controlled conditions. Hence, we open the chance to extended the applicability of noise as a resource for
quantum computational algorithms in general. We hope that our point of view will contribute to more uses of available NISQ technology.

\appendix

\section{Derivation of \texorpdfstring{$\Lam(\theta;\gamma)$}{Lambda(theta,gamma)}}
\label{app:jacobian_theta}

Let us consider the vector
\begin{equation}
\mathbf r'(\theta,\phi)=
\begin{pmatrix}
\sqrt{A}\,\sin\theta\cos\phi\\
\sqrt{A}\,\sin\theta\sin\phi\\
A\cos\theta+\gamma
\end{pmatrix},
\end{equation}
and define
\begin{equation}
N(\theta):=\|\mathbf r'(\theta,\phi)\|.
\end{equation}
Since $N(\theta)$ is the Euclidean norm, we have
\begin{align}
N(\theta)^2
&=
\left(\sqrt{A}\sin\theta\cos\phi\right)^2
+\left(\sqrt{A}\sin\theta\sin\phi\right)^2
+\left(A\cos\theta+\gamma\right)^2 \nonumber\\
&=
A\sin^2\theta\left(\cos^2\phi+\sin^2\phi\right)
+\left(A\cos\theta+\gamma\right)^2 \nonumber\\
&=
A\sin^2\theta+\left(A\cos\theta+\gamma\right)^2.
\label{eq:N2_def}
\end{align}

The angular map is given by
\begin{equation}
\cos\theta'=\frac{A\cos\theta+\gamma}{N(\theta)},
\qquad
\phi'=\phi,
\label{eq:map_def}
\end{equation}
or equivalently,
\begin{equation}
\theta'=\arccos\!\left(\frac{A\cos\theta+\gamma}{N(\theta)}\right).
\end{equation}

\subsection{Jacobian of the map}
Since $\phi'=\phi$ and $\theta'$ does not depend on $\phi$, the Jacobian is triangular:
\begin{equation}
J(\theta,\phi)=
\begin{pmatrix}
\dfrac{\partial\theta'}{\partial\theta} & \dfrac{\partial\theta'}{\partial\phi}\\[6pt]
\dfrac{\partial\phi'}{\partial\theta} & \dfrac{\partial\phi'}{\partial\phi}
\end{pmatrix}
=
\begin{pmatrix}
\dfrac{d\theta'}{d\theta} & 0\\
0 & 1
\end{pmatrix}.
\label{eq:jacobian_triangular}
\end{equation}
Therefore, the only nontrivial term is $d\theta'/d\theta$.

\subsection{Closed-form derivation of $\dfrac{d\theta'}{d\theta}$}

We define
\begin{equation}
p(\theta):=A\cos\theta+\gamma,
\qquad
u(\theta):=\frac{p(\theta)}{N(\theta)}.
\label{eq:p_u_def}
\end{equation}
Then, by \eqref{eq:map_def}, we have $\cos\theta'=u(\theta)$.
Differentiating implicitly with respect to $\theta$:
\begin{equation}
\frac{d}{d\theta}\cos\theta'
=
-\sin\theta'\,\frac{d\theta'}{d\theta}
=
\frac{du}{d\theta}.
\end{equation}
Hence,
\begin{equation}
\frac{d\theta'}{d\theta}
=
-\frac{u'(\theta)}{\sin\theta'}.
\label{eq:dtheta_general}
\end{equation}

\paragraph{(i) Expression of $\sin\theta'$ in terms of $\theta$.}
Using $\cos\theta'=u$,
\begin{align}
\sin^2\theta'
&=
1-\cos^2\theta'
=
1-u(\theta)^2
=
1-\frac{p(\theta)^2}{N(\theta)^2} \nonumber\\
&=
\frac{N(\theta)^2-p(\theta)^2}{N(\theta)^2}.
\end{align}
From \eqref{eq:N2_def} we have $N(\theta)^2-p(\theta)^2=A\sin^2\theta$, hence
\begin{equation}
\sin^2\theta'
=
\frac{A\sin^2\theta}{N(\theta)^2},
\qquad
\Rightarrow
\qquad
\sin\theta'=\frac{\sqrt{A}\,\sin\theta}{N(\theta)}.
\label{eq:sin_theta_prime}
\end{equation}
(For the usual range $0\le\theta\le\pi$ we take $\sin\theta\ge 0$.)

\paragraph{(ii) Computation of $u'(\theta)$.}
Since $u=p/N$, by the quotient rule:
\begin{equation}
u'(\theta)=\frac{p'(\theta)\,N(\theta)-p(\theta)\,N'(\theta)}{N(\theta)^2}.
\label{eq:u_prime}
\end{equation}
Moreover,
\begin{equation}
p'(\theta)=\frac{d}{d\theta}\left(A\cos\theta+\gamma\right)=-A\sin\theta.
\label{eq:p_prime}
\end{equation}

Substituting \eqref{eq:u_prime} and \eqref{eq:sin_theta_prime} into \eqref{eq:dtheta_general}:
\begin{align}
\frac{d\theta'}{d\theta}
=
-\frac{\dfrac{p'N-pN'}{N^2}}{\dfrac{\sqrt{A}\sin\theta}{N}}
=
-\frac{p'N-pN'}{N\sqrt{A}\sin\theta}
=\\
-\frac{p'}{\sqrt{A}\sin\theta}
+\frac{p}{\sqrt{A}\sin\theta}\frac{N'}{N}.
\label{eq:dtheta_split}
\end{align}
With \eqref{eq:p_prime} we obtain
\begin{equation}
-\frac{p'}{\sqrt{A}\sin\theta}
=
-\frac{-A\sin\theta}{\sqrt{A}\sin\theta}
=
\sqrt{A},
\end{equation}
and therefore
\begin{equation}
\frac{d\theta'}{d\theta}
=
\sqrt{A}
+\frac{p}{\sqrt{A}\sin\theta}\frac{N'}{N}.
\label{eq:dtheta_mid}
\end{equation}

\paragraph{(iii) Computation of $\dfrac{N'}{N}$.}
We differentiate \eqref{eq:N2_def}:
\begin{equation}
N(\theta)^2=A\sin^2\theta+p(\theta)^2.
\end{equation}
Then
\begin{align}
2NN'
&=
2A\sin\theta\cos\theta+2p\,p' \nonumber\\
\Rightarrow\quad
\frac{N'}{N}
&=
\frac{A\sin\theta\cos\theta+p\,p'}{N^2}.
\label{eq:Nprime_over_N}
\end{align}
Using $p'=-A\sin\theta$ in \eqref{eq:Nprime_over_N}:
\begin{equation}
\frac{N'}{N}
=
\frac{A\sin\theta\cos\theta-Ap\sin\theta}{N^2}
=
\frac{A\sin\theta(\cos\theta-p)}{N^2}.
\label{eq:Nprime_over_N_simplified}
\end{equation}

Substituting \eqref{eq:Nprime_over_N_simplified} into \eqref{eq:dtheta_mid}:
\begin{align}
\frac{d\theta'}{d\theta}
&=
\sqrt{A}
+
\frac{p}{\sqrt{A}\sin\theta}\cdot
\frac{A\sin\theta(\cos\theta-p)}{N^2} \nonumber\\
&=
\sqrt{A}
+
\sqrt{A}\,\frac{p(\cos\theta-p)}{N^2}
=
\sqrt{A}\left(1+\frac{p\cos\theta-p^2}{N^2}\right) \nonumber\\
&=
\sqrt{A}\,\frac{N^2+p\cos\theta-p^2}{N^2}.
\label{eq:dtheta_before_final_simplify}
\end{align}
Using $N^2=A\sin^2\theta+p^2$ in the numerator of \eqref{eq:dtheta_before_final_simplify}:
\begin{align}
N^2+p\cos\theta-p^2
&=
A\sin^2\theta+p^2+p\cos\theta-p^2
=
A\sin^2\theta+p\cos\theta \nonumber\\
&=
A(1-\cos^2\theta)+(A\cos\theta+\gamma)\cos\theta \nonumber\\
&=
A-A\cos^2\theta+A\cos^2\theta+\gamma\cos\theta
=
A+\gamma\cos\theta.
\end{align}
Therefore,
\begin{equation}
\boxed{
\frac{d\theta'}{d\theta}
=
\sqrt{A}\;
\frac{A+\gamma\cos\theta}{N(\theta)^2}
}.
\label{eq:dtheta_general_closed}
\end{equation}

\subsection{Case $A=1-\gamma$}
In the case used in the main text, $A=1-\gamma$. Moreover, from the norm calculation (see \eqref{eq:N2_def} and the corresponding
simplification) one has
\begin{equation}
N(\theta)^2
=
1-\gamma(1-\gamma)\bigl(1-\cos\theta\bigr)^2.
\label{eq:N2_final}
\end{equation}
The numerator of \eqref{eq:dtheta_general_closed} becomes
\begin{equation}
A+\gamma\cos\theta=(1-\gamma)+\gamma\cos\theta
=
1-\gamma\bigl(1-\cos\theta\bigr).
\end{equation}
Substituting into \eqref{eq:dtheta_general_closed} and using \eqref{eq:N2_final}:
\begin{equation}
\boxed{
\frac{d\theta'}{d\theta}
=
\sqrt{1-\gamma}\;
\frac{1-\gamma\bigl(1-\cos\theta\bigr)}
{1-\gamma(1-\gamma)\bigl(1-\cos\theta\bigr)^2}
}.
\label{eq:dtheta_final}
\end{equation}

Finally, the Jacobian of the map \eqref{eq:map_def} can be written as
\begin{equation}
\boxed{
J(\theta,\phi)=
\begin{pmatrix}
\dfrac{d\theta'}{d\theta} & 0\\
0 & 1
\end{pmatrix}
}
\qquad\text{with}\quad
\frac{d\theta'}{d\theta}\ \text{given by \eqref{eq:dtheta_final}.}
\end{equation}


\end{document}